\documentclass[a4paper,twocolumn]{revtex4-1}
\usepackage[utf8]{inputenc}
\usepackage[portuguese]{babel}
\usepackage{amsmath}
\usepackage{amsfonts}
\usepackage{amssymb}
\usepackage[left=2cm,right=2cm,top=2cm,bottom=2cm]{geometry}

\begin{document}
\title{The overestimated potential of solar energy to mitigate climate change}
\date{\today}
\author{Marcos Paulo Belançon}
\email{marcosbelancon@utfpr.edu.br}
\affiliation{Programa de Pós-graduação em Processos Químicos e Bioquímicos\\Universidade Tecnológica Federal do Paraná\\CEP 85503-390, Pato Branco, Paraná, Brasil}

\begin{abstract}
Many aspects of solar energy and policies to tackle the energy transition have been neglected. Even though the earth is plenty of sun energy, our planet is not plenty of resources to transform that energy into electricity. This is a case between many others where an strongly optimistic bias is shadowing the white elephant in the room.
\end{abstract}

\maketitle

In the defense of Photovoltaics (PV) Creutzig et al published their view on ``The underestimated potential of solar energy to mitigate climate change''\cite{Creutzig2017}. They wrote, for example: ``Direct solar energy has a technical potential of 1,500–50,000 EJ per year, exceeding the projected global primary energy demand of about 1,000 EJ per year in 2050''.

We should pay attention to what exactlly means ``technical''in such sentence. If technically available means achievable with today's technology, the statement is false. On the other hand, if it is interpreted with the belief that science will figure out a solution the statement may be true. In a previous work\cite{Belancon2017d} we have presented a point of view where this optimism is a threat, which may be delaying the real discussion that we should made. 

The sun provides virtually ``limitless energy'', and the most efficient devices we have to use this energy are solar water heaters; those devices may reach efficiencies as high as 70\% at full sun. Thermal energy, however, is a ``low quality'' energy that can not be used to cover all of our needs. By this way, we aim to convert more sunlight into electricity. 

The PV market is by far dominated by Silicon, a technology discovered half a century ago. It is true that Silicon PV is quite cheap today, however ``symbolic human prices'' measured in dollars may not reflect the ``real natural prices''. We are consuming 25 kg of Silver per MWp of Silicon PV built\cite{silver}. This means that with the today's production rate running around $~100GWp$ we are consuming 2500 metric tons of Silver per year; that is 10\% of global extraction of Silver.

This PV production can guarantee only $2500GWp$ of PV installed, because in 25 years from now we will be replacing the PV's that are new today. If we consider a capacity factor of about 20\%, something like $500GW$, or about 5\% of mankind's primary energy is all the solar electricity we are going to have in 2040 if we do not expand the production. Even though many optimists will say that this is exactly what is going to happen, one may point that it is not so clear that the industry will reduce even more the Silver consumption; the learning curve of Silicon PV industry is already mature, and there is not much space for improve production efficiencies and reduce prices. Other technologies such as thin film PV's rely on Cadmium, Tellurium, Indium and Selenium, all of those will limit the production rate of thin films far below the Silicon PV's level. 

One may see in the news that many companies in Japan, Europe and North America are declaring insolvency and/or loosing stock value due cheaper China PV's. There are even some doubt about if the Chinese companies are really making money; the hyphothesis that the demand is lower than the supply of PV's can be perfectly sustained for now.

I do like solar energy, but we don't have scientifical evidence to sustain that our civilization lifestyle will be saved by Photovoltaics. Many are concerned about efficiencies of solar cells, but we do not have a problem with that. An average brazilian house may produce its own electricity with about $15m^2$ of commercial Silicon PV; on the other hand, if we want the same amount of energy from biomass, for example, a hundred times more area should be necessary. Our technologies are already far more efficient than photosynthesis.

One may ask, then, why Brazil has so much biomass and little PV's? The Brazilian geography makes possible that huge areas of land can be used for Sugar Cane production; Brazil's land used to produce sugar cane reached 90.000$km^2$ this year, which is about the size of Portugal. Brazil's primary energy is 41\% renewable\cite{ben2017}, and it is oftenly interpreted that Brazil achieve this because of its great hydroeletric power. However, Sugar Cane produces 16.7\% of the primary energy while hydro produces 11\%. All biomass combined produces twice more energy than hydro, and with Sugar Cane an efficient liquid fuel is obtained. Ethanol and electricity from Sugar Cane in Brazil accounts for 580 $TWh$\cite{ben2017} of primary energy; electricity from PVs in Germany in 2016 accounted for only 38 $TWh$\cite{energygermany}. The secrets of Brazil energy mix are 1) location, not far from equator, 2) half of the world population density, 3) almost no coal resources available and 4) consumption per capita equals to the world average.

Many will also point out that governments should tax fossil fuels, meanwhile the United States have choosen the opposite direction during the Obama administration. They reduced the direct emission of Green House Gases (GHG) by replacing Coal by Shale Gas to generate electricity\cite{Obama2017}. Shale oil and Canadian oil sands provided a way to avoid middle east fossil fuels; by this way an ``energy war'' between OPEC and North America pushed oil prices down. Those policies have made gas electricity so cheap in US that some nuclear plants were shutdown this year, what in turn will reduce the effectiveness of Shale gas in reduce GHG emissions further. And Shale has already give signals that if we stop drilling new wells the output may go down pretty fast\cite{Buurma2017}. The optimism bias of those standing for PV's are analogous to that of Obama's view. Both are looking to a small scale of the challenge and forgetting many aspects when claiming how effective Shale or PV's may be to solve our energy dilemma.

There are no alternative to fossil fuels able to reach the world scale because we can not produce sugar cane everywhere, neither produce as much PV's as we would like to. Many other questions such as storage are important but it does not come first than our capability to produce the PV's. By one way or another, the age of rapid growth in energy supply seems to be going to an end, because the energy return on energy invested is decreasing\cite{Taylor2016}. We are moving towards low quality and environmentally costly fossil fuels at the same pace we try to build a renewable energy infrastructure that relies on complex technologies that demand increasing extraction rate of finite scarse minerals. 

If we let apart the faith that science will continuosly improve technologies and innovate making our civilization run as a ``perpetual motion machine'', one may state that the second law of thermodynamics guarantees that natural resources, such as oil or silver, will become less abundant and more costly to extract. By this way, in the short term some companies and countries will go bankrupty due the low prices, as we have seen, and next the cost of fossil fuels, PV's and every energy based on scarse resources will become more expensive. 

The belief in endless growth of consumption that have been central in our civilization makes our behavior be very similar to that of a microbial culture. It is true that overpopulation puts pressure in our ecosystem, however, it is also true that during the last century our consumption per person has grown faster than the population. The energy transition is a great goal for our civilization, but it may be useless if we do not rethink our civilization itself.
\bibliography{/home/mbelancon/Documentos/library}

\end{document}